\documentclass[12pt]{article}
\usepackage{graphicx}

\newcommand\pubnumber{DPF2013-49}
\newcommand\pubdate{\today}

\def\place{on behalf of the LHCb Collaboration\\
CERN, CH-1211, Geneva 23, SWITZERLAND}

\def\Title#1{\begin{center} {\Large #1 } \end{center}}
\def\Author#1{\begin{center}{ \sc #1} \end{center}}
\def\Address#1{\begin{center}{ \it #1} \end{center}}

\newcommand\pubblock{\rightline{\begin{tabular}{l} \pubnumber\\
         \pubdate  \end{tabular}}}
\newenvironment{Abstract}{\begin{quotation}  }{\end{quotation}}
\newenvironment{Presented}{\begin{quotation} \begin{center} 
             PRESENTED AT\end{center}\bigskip 
      \begin{center}\begin{large}}{\end{large}\end{center} \end{quotation}}

\def\beq{\begin{equation}}
\def\eeq#1{\label{#1}\end{equation}}
\def\eeqn{\end{equation}}

\def\beqa{\begin{eqnarray}}
\def\eeqa#1{\label{#1}\end{eqnarray}}
\def\eeqan{\end{eqnarray}}

\let\bar=\overbar

\def\Dslash{\not{\hbox{\kern-4pt $D$}}}
\def\dslash{\not{\hbox{\kern-2pt $\del$}}}

\def\msb{{\bar{\ssstyle M \kern -1pt S}}}

\begin{document}
\begin{titlepage}
\pubblock

\vfill
\Title{The LHCb Upgrade}
\vfill
\Author{ Federico Alessio}
\Address{\place}
\vfill
\begin{Abstract}
The LHCb experiment is designed to perform high-precision measurements of CP violation and search for New Physics using the enormous flux involving beauty and charm quarks produced at the LHC. The operation and the results obtained from the data collected in 2010 and 2011 demonstrate that the detector is robust and functioning very well. However, the limit of 1~fb$^{-1}$ of data per nominal year cannot be overcome without improving the detector. We therefore plan for an upgraded spectrometer by 2018 with a 40~MHz readout and a much more flexible software-based triggering system that will increase the data rate as well as the efficiency specially in the hadronic channels. Here we present the LHCb detector upgrade plans, based on the Letter of Intent and Framework Technical Design Report.
\end{Abstract}
\vfill
\begin{Presented}
DPF 2013\\
The Meeting of the American Physical Society\\
Division of Particles and Fields\\
Santa Cruz, California, August 13--17, 2013\\
\end{Presented}
\vfill
\end{titlepage}
\def\thefootnote{\fnsymbol{footnote}}
\setcounter{footnote}{0}

\section{Introduction}

The LHCb experiment \cite{LHCb} is a high-precision experiment at the LHC devoted to the search for New Physics by precisely measuring its effects in CP violation and rare decays. By applying an indirect approach, LHCb is able to probe effects which are strongly suppressed by the Standard Model, such as those mediated by loop diagrams and involving flavor changing neutral currents. This is a powerful approach as it allows accessing larger mass scales than those reachable by direct searches at the other LHC experiments. These GPDs apply a more direct approach and are bounded to the center-of-mass energy scale accessible at the LHC.

\begin{figure}[htb]
\centering
\includegraphics[height=2.8in]{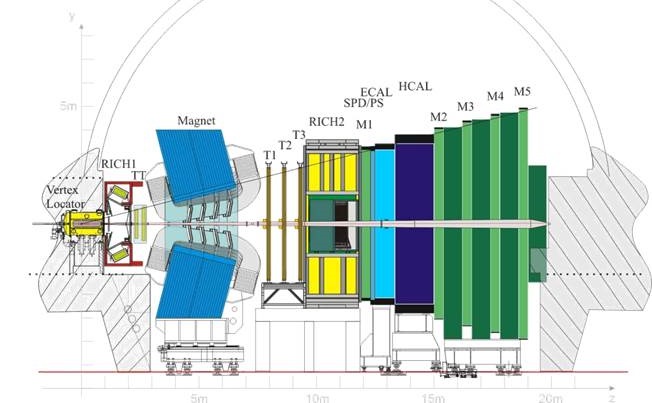}
\caption{Layout of the LHCb detector.}
\label{fig:detector}
\end{figure}

\begin{figure}[htb]
\centering
\includegraphics[height=1.8in]{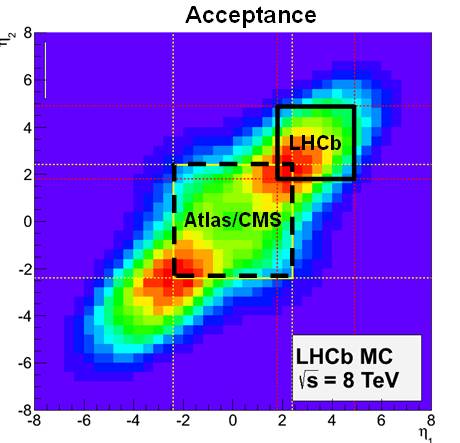}
\caption{Illustration of the acceptance complementarity as a function of pseudo-rapidity between experiments at the LHC.}
\label{fig:rapidity}
\end{figure}

In the proton-proton collision mode, the LHC is to a large extent a heavy flavor factory producing over 100.000~b\={b}-pairs every second at the nominal LHCb design luminosity of 2x10$^{32}$~cm$^{-2}$~s$^{-1}$. Given that  b\={b}-pairs are predominantly produced in the forward or backward direction, the LHCb detector was designed as a forward spectrometer (Figure~\ref{fig:detector}) with the detector elements installed along the main LHC beamline, covering a pseudo-rapidity range of 2 $< \eta <$ 5 well complementing the other LHC detectors ranges. This is illustrated in Figure~\ref{fig:rapidity}. 

A silicon-strip Vertex Locator (VELO) detector provides excellent impact parameter resolution of 20~$\mu$m for high-p$_{T}$ tracks and proper time resolution of 45~fs for $B_{s} \rightarrow J/\psi\phi$ and for $B_{s} \rightarrow D_{s}\pi$. Combined with large areas of silicon (Tracker Turicensis, TT and Inner Tracker, IT) and straw tubes (Outer Tracker, OT) tracking stations located around a 4~Tm spectrometer magnet, they provide a momentum resolution of $\Delta$p/p = 0.4$\%$ at 5~GeV/c to 0.6$\%$ at 100~GeV/c with a track reconstruction efficiency of more than 96$\%$ for long tracks. Two Ring-Imaging Cherenkov (RICH) detectors are used to identify charged hadrons, with a kaon ID efficiency of about 95$\%$ for about 5$\%$ $\pi \rightarrow K$ mis-ID probability. Moreover, an Electromagnetic Calorimeter (ECAL) identifies photons and electrons with an efficiency of about 90$\%$ for about 5$\%$ $e \rightarrow h$ mis-ID probability and an Hadronic Calorimeter (HCAL) is used to identify hadrons. Lastly, further downstream, a series of alternating layers of iron and Multi-Wire Proportional Chambers (MWPC) and GEMs modules are used to distinguish muons from hadrons with an efficiency of about 97$\%$ for 1-3$\%$ $\pi \rightarrow \mu$ mis-ID probability. The Calorimeter and the Muon detectors are inputs to the first-level hardware trigger of the readout electronics of the detector.

Furthermore, LHCb proved excellent performance in terms of data taking \cite{Operation}, operating with close to 99$\%$ of working detector channels and recording more than 90$\%$ of the data delivered by the LHC adding up to a total of about 3~fb$^{-1}$ of data saved to disk over the period 2010-2012. More than 99$\%$ of the data recorded online was already good for offline analyses, thanks to the excellent background conditions at the LHC as well. This corresponds to about 60x10$^{11}$~c\={c}-pairs and 26x10$^{10}$~b\={b}-pairs on disk. The experiment was able to record such a high number of events by exceeding its nominal design running conditions by more than a factor two in terms of instantaneous luminosity by running at a leveled \cite{Leveling} instantaneous luminosity of 4x10$^{32}$~cm$^{-2}$~s$^{-1}$ and by more than a factor four in terms of average number of interactions per crossing - $\mu$ = 1.5 instead of the nominal $\mu$ = 0.4. LHCb proved itself to be the Forward General-Purpose Detector at the LHC well complementing the physics results from ATLAS and CMS. 

\section{Motivations for an upgraded experiment}
\begin{figure}[htb]
\centering
\includegraphics[height=2.5in]{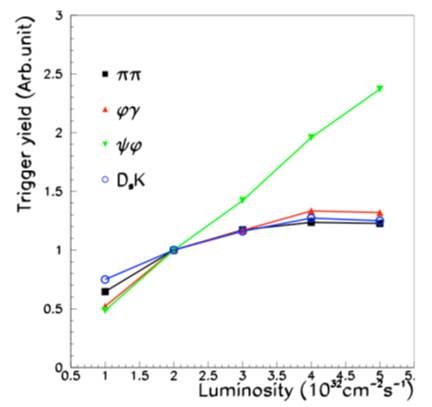}
\caption{Physics yields as a function of the instantaneous luminosity for some hadronic and muonic channels.}
\label{fig:trigger}
\end{figure}

After the LHC Long Shutdown I (2013-2014), the LHCb experiment will keep recording data, improving the statistics of its measurements. It is expected to take in excess of about 5~fb$^{-1}$ by 2018, by recording data at a constant luminosity of 4x10$^{32}$~cm$^{-2}$~s$^{-1}$ throughout the period 2015-2018. Hence, the total LHCb dataset will add up to about 8~fb$^{-1}$ by 2018 as it stands at the writing of this paper. During the same period, however, the LHC accelerator will increase the total center-of-mass energy to 13~TeV and it will change the bunch spacing from the current 50~ns to the nominal 25~ns. This is expected to allow the LHC to reach its nominal designed luminosities in ATLAS and CMS of about 2x10$^{34}$~cm$^{-2}$~s$^{-1}$ while keeping the complexity of events low as $\mu$ will decrease. In LHCb, this means that the amount of beauty and charm decays generated will essentially double, while reducing the complexity of events by a factor two. As the current LHCb detector is already performing very well under conditions which are harsher than the designed ones and the LHC accelerator will keep improving its performance, the prospect to augment the physics yield in the LHCb dataset seems very attractive, following the improved performance of the LHC. However, the LHCb detector is limited by design in terms of data bandwidth - 1~MHz instead of the LHC bunch crossing frequency of 40~MHz - and physics yield for hadronic channels at the hardware trigger, therefore it will not be possible to increase the physics yield during the period 2015-2018. Figure~\ref{fig:trigger} illustrates the physics yield at the LHCb first-level hardware trigger as a function of the instantaneous luminosity. At a luminosity of 4x10$^{32}$~cm$^{-2}$~s$^{-1}$ the physics yield for hadronic channels is essentially half of the one for muonic channels. A Letter Of Intent \cite{LOI} and Framework TDR \cite{FTDR} document the plans for an upgraded detector which will enable LHCb to increase the yield in the decays with muon by a factor of 10, the yield for hadronic channels by a factor 20 and to collect at least 50~fb$^{-1}$ at a leveled constant luminosity of 1-2x10$^{33}$~cm$^{-2}$~s$^{-1}$. This corresponds to ten times the current design luminosity.

By increasing the physics yields, the LHCb will perform measurements beyond Flavor Physics. Many of the current exploration studies will become precision studies, enabling LHCb to reach a sensitivity comparable to the theory's one. A more exhaustive list of measurements can be found in \cite{FTDR}, however the main highlights are that LHCb will help constraining the CKM $\gamma$ angle to $<$ 1$\%$ and to measure the BR($B_{s} \rightarrow \mu^{+}\mu^{-})$ down to about 10$\%$ of Standard Model.

\section{Strategy for the upgrade of the experiment}
In order to remove the main design limitations of the current LHCb detector, the strategy for the upgrade of the LHCb experiment consists of ultimately removing the first-level hardware trigger entirely, hence to run the detector fully trigger-less. Currently, the Calorimeters and Muon detectors provide p$_{T}$, E$_{T}$ and muon information to the first-level hardware trigger. Based on that information, the rate of accepted events is reduced from the LHC bunch crossing frequency of 40~MHz to about 1~MHz. Events are then built and transmitted through a 1~Tb/s readout network to a processing farm consisting of about 30000 cores. A software trigger running on each core of the processing farm analyses the events and overall selects about 5~kHz of them to be saved to disk. 

By removing the first-level hardware trigger, LHC events are recorded and transmitted from the Front-End electronics to the readout network at the full LHC bunch crossing rate of 40~MHz, resulting in a multi-Tb/s network. All events will therefore be available at the processing farm where a fully flexible software trigger will perform selection on events, with an overall output of about 20~kHz of events to disk. This will allow maximizing signal efficiencies at high event rates.

\begin{figure}[htb]
\centering
\includegraphics[height=2.8in]{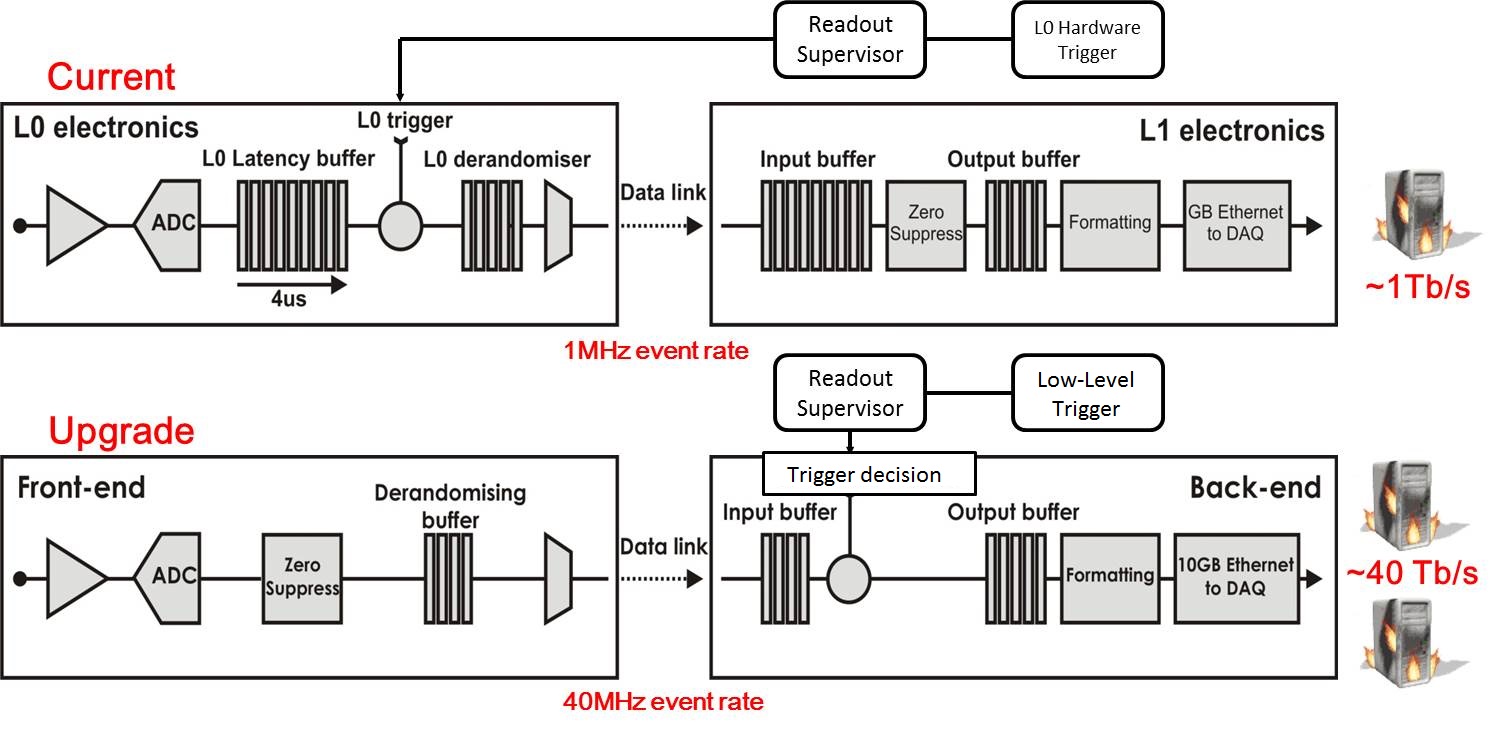}
\caption{The upgrade LHCb trigger-less readout architecture.}
\label{fig:readout}
\end{figure}

The direct consequences of this approach are that some of the LHCb sub-detectors will need to be redesigned to cope with an average luminosity of up to 2x10$^{33}$~cm$^{-2}$~s$^{-1}$ and to be equipped with new trigger-less Front-End electronics, and the entire readout architecture must be redesigned in order to cope with the upgraded multi-Tb/s bandwidth and a full 40~MHz dataflow.

A completely new VELO is being developed. It is based on silicon-pixels with microchannel cooling, with reduced budget material, higher radiation hardness and improved resolution to cope with higher multiplicities. The new detector is expected to maintain at least the same performance as the current VELO, but in ten times harsher conditions. It is also under study the possibility to reduce the inner aperture of the VELO from the current 5.5~mm to a possible 3.5~mm. This will improve the impact parameter resolution. 

The present tracking system will be entirely upgraded as well. Various options are still under study for the upstream (before the LHCb dipole) and downstream (after) stations. For the upstream stations, a new Upstream Tracker (UT) is being developed, based on silicon-strips. However, it will have a reduced thickness, finer granularity, an improved coverage and much less material budget ($<$~5$\%$~X$_{0}$). It was shown that by combining the new VELO and UT, the momentum resolution will improve by about 5$\%$ and reduce the ghost rate, even in harsher conditions. For the downstream stations, a new conceptual detector is being developed, based on scintillating fibers (Sci-Fi). This detector will replace the region occupied by the current IT and OT. It will be formed by 12 layers of 2x2.5~m long scintillating fibers, laid down as mat. Fibers will have 250~$\mu$m diameter and they will be read out with Si-PM cooled down to well below -30$^\circ$C.

The RICH detector will be upgraded as well. It was chosen to pursue a light upgrade, in which the current aerogel radiator is removed and the optics will be redesigned to compensate for the increased occupancy. The current Hybrid Photo Detectors (HPD) are replaced by Multi Anode PMTs (MaPMT) and the Front-End electronics will have to be redeveloped as well in order to be compatible with the fully trigger-less readout architecture.

The Calorimeters and Muon detectors will only replace their Front-End electronics to be compatible with the fully trigger-less readout architecture, while maintaining the current detector technology. 

The entire readout architecture will need to be upgraded in order to cope with the full 40~MHz dataflow \cite{Readout}. Figure~\ref{fig:readout} schematically illustrates the main differences between the current readout architecture and the proposed upgraded one. Each LHCb sub-detector will be equipped with trigger-less Front-End (FE) electronics, able to record, compress and transmit data packets to the Back-End (BE) electronics. No triggers are directly transmitted to the FE, therefore the FE must be able to compress data before transmit it through the link. Moreover, another important requirement is that the FE must be able to use data links bandwidth efficiently in order to globally reduce the number of links from the detector to the BE. In this sense, compression algorithms and complex packing mechanisms are being developed to meet the requirements. Clock, fast commands and slow control are transmitted to the FE by sharing the same compact links and by profiting from the common CERN development on the GigaBit Transceiver (GBT) \cite{GBT}. 

Regarding the BE, the current baseline approach envisages to use of a very compact, high density, FPGA-based ATCA \cite{ATCA} readout board \cite{TELL40}. This board will be able to handle a throughput of $>$~0.5~Tb/s alone, by profiting from an extremely high link density on board. The connection to the processing farm will be via 10~Gigabit Ethernet links and big and powerful switches. The board is so generic that can be used also for the fast control, clock distribution and slow control distribution. Another more innovative approach takes ideas from the current data-centers being used in many commercial environment. In this solution, a more compact network could be envisaged in which each processing node is also a host for a PCIe NIC readout board able to handle up to 100~Gb/s. This board will actually act as a readout board and the event building is performed through a series of cheaper and commercially available switches in a more compact way. Disks and monitoring computers will also be within the same network, composing what could be seen as a readout box. An R$\&$D phase is ongoing in order to evaluate the feasibility of each solution.

It is to note that even though the ultimate goal is to entirely remove the first-level hardware trigger, a similar version of the current hardware trigger will be maintained in order to help staging the installation of the processing farm if needed. The trigger is commonly referred to as Low-Level-Trigger (LLT) and it will be used to tune the readout rate between the current 1~MHz to the maximum 40~MHz.

\section{Conclusion}
The LHCb experiment has proposed an upgrade of its detector and readout electronics in order to increase its physics yields by a factor 10 in muonic channels, a factor 20 in hadronic channels and a factor 10 in integrated luminosity by removing the first-level hardware trigger. The plan to upgrade the detector in 2018 consists of replacing some sub-detectors to newest technologies and to equip them with trigger-less Front-End electronics, while retaining the current performance. Moreover, in order to cope with a multi-Tb/s readout network, the entire readout architecture has been redesigned and redeveloped. All LHC events will be recorded at the LHC bunch crossing frequency and made available to the processing farm in order to maximize signal efficiencies while increasing the readout rate. 

The installation and commissioning of the upgraded detector and readout system is planned for the LHC Long Shutdown II (2018-2019) over a period of about 18 months. After that, the LHCb experiment is expected to collect at least 50~fb$^{-1}$ of at a leveled luminosity of 1-2x10$^{33}$~cm$^{-2}$~s$^{-1}$. This is expected to increase dramatically the LHCb sensitivity beyond Flavor Physics and in the search for New Physics. CERN has fully endorsed the upgrade by officially approving it during the course of 2012.

\end{document}